\documentclass[ALICE,manyauthors]{cernphprep}

\usepackage[comma,square,numbers,sort&compress]{natbib}

\usepackage[T1]{fontenc} 
\usepackage[utf8]{inputenc}
\usepackage[english]{babel}

\usepackage{amssymb}
\usepackage{hyperref}
\usepackage{lineno}
\usepackage[amssymb]{SIunits}
\usepackage{multicol}
\usepackage{multirow}
\usepackage{units}
\usepackage{subcaption}
\usepackage{graphics}
\usepackage{textcomp} 

\newcommand{\pp}           {pp}

\newcommand{\incldNdeta}       {\ensuremath{\mathrm{d}N_\mathrm{ch}^\mathrm{incl.}/\mathrm{d}\eta}}

\newcommand{\s}            {\ensuremath{\sqrt{s}}}
\newcommand{\pt}           {\ensuremath{p_{\rm T}}}
\newcommand{\pT}           {\ensuremath{p_{\rm T}}}

\newcommand{\inelg}     {\ensuremath{\mathrm{INEL}_{>0}}}
\newcommand{\multc} {\ensuremath{\Delta\sigma/\sigma_{\rm MB_{AND>0}}}}
\newcommand{\multcs} {\ensuremath{\sigma/\sigma_{\rm MB_{AND>0}}}}
\newcommand{\multcc} {\ensuremath{\Delta\sigma/\sigma_{\rm INEL_{>0}}}}

\newcommand{\mbandg} {\ensuremath{\mathrm{ MB_{AND>0}}}}
\newcommand{\mband} {$\rm{MB}_{AND}$}

\newcommand{\dndeta}{\ensuremath{{\rm d}N_{\rm ch}/{\rm d}\eta}\xspace}

\newcommand{\avdndeta}{\ensuremath{\langle \dndeta \rangle}}
\newcommand{\navdndeta}{\ensuremath{\langle \dndeta \rangle / \langle \dndeta \rangle_\mathrm{incl.}}}
\newcommand{\zvtx}{\ensuremath{z_\mathrm{vtx}}}

\begin{document}

\begin{titlepage}
\PHyear{2020}        
\PHnumber{170}      
\PHdate{07 September}  
%

\title{Pseudorapidity distributions of charged particles as a function of mid- and forward rapidity multiplicities in pp collisions at \s~=~5.02, 7 and \unit[13]{TeV} }
\ShortTitle{Multiplicity dependence study of the pseudorapidity density}   

\Collaboration{ALICE Collaboration\thanks{See Appendix~\ref{app:collab} for the list of collaboration members}}
\ShortAuthor{ALICE Collaboration} 

\begin{abstract}
  The multiplicity dependence of the pseudorapidity density of charged particles in proton--proton (\pp) collisions at centre-of-mass energies $\sqrt{s}$~=~5.02, 7 and \unit[13]{TeV} measured by ALICE is reported. The analysis relies on track segments measured in the midrapidity range ($|\eta| < 1.5$). Results are presented for inelastic events having at least one charged particle produced in the pseudorapidity interval $|\eta|<1$.  
  The multiplicity dependence of the pseudorapidity density of charged particles is measured with mid- and forward rapidity multiplicity estimators, the latter being less affected by autocorrelations. A detailed comparison with predictions from the PYTHIA 8 and EPOS LHC event generators is also presented. 
  The results can be used to constrain models for particle production as a function of multiplicity in pp collisions.

  \end{abstract}
  
  \end{titlepage}
  \setcounter{page}{2}

  \section{Introduction}
  
  The study of high-multiplicity events in proton--proton (pp) and proton--nucleus high-energy collisions reveals striking similarities with respect to the observations made for larger systems like a nucleus--nucleus collision, which are interpreted in terms of the creation of a strongly-interacting, fluid-like QCD medium: the quark--gluon plasma (QGP). The ridge structure arising from long-range azimuthal correlations observed in pp data~\cite{Khachatryan:2010gv,Khachatryan:2015lva,Aad:2015gqa} is also found in p--Pb collisions~\cite{Abelev:2012ola,Aad:2012gla,CMS:2012qk,Aad:2014lta}, where the presence of double-ridge structures is reported~\cite{Abelev:2012ola}. More recently, an ALICE measurement reported an enhancement in the relative production of (multi-) strange particles with respect to primary charged particles as a function of multiplicity in pp collisions~\cite{ALICE:2017jyt}. This suggests that some observables related to the QGP formation might be driven just by the multiplicity regardless of collision systems at LHC energies.
  
  In \pp\ and p--Pb collisions, the selection of events with large hadronic final-state multiplicities biases the sample towards a large average number of Multiple Parton Interactions (MPIs) at the LHC~\cite{Abelev:2013sqa,Abelev:2014mva}. 
  In the description provided by the colour reconnection (CR) mechanism~\cite{Christiansen:2015yqa,Ortiz:2013yxa}, 
  CR in MPIs is expected to be particularly pronounced at high multiplicity. The effects of prominent CR at high multiplicity are supposed to account for basic observables like the correlations between the average momentum and the multiplicity of charged particles~\cite{Aad:2010ac} as well as for the shape of their pseudorapidity distribution~\cite{Gieseke:2012ft}. Indeed, the transverse momentum ($\pT$) spectra of charged particles at high multiplicity~\cite{Abelev:2013bla, Aaboud:2016itf} can be attributed, in pp collisions, to a CR mechanism, while until now no multiplicity dependence study of charged particle pseudorapidity density has been published.

  This document provides a large set of charged-particle multiplicity density measurements as a function of event multiplicity in pp collisions at different centre-of-mass energies. This work could shed light on the phenomenon of MPIs that is a key ingredient of models attempting to describe large-multiplicity events. In any collision system, the event-averaged pseudorapidity density of primary charged particles~\cite{ALICE-PUBLIC-2017-005}, \dndeta, is a key observable characterising the global properties of the collision. Especially in \pp\ interactions, the \dndeta\ is described  
  by the combination of the perturbative hard partonic processes and the underlying event~\cite{ALICE:2011ac,Acharya:2019nqn}. The underlying event includes various phenomena like initial- and final-state radiation, colour-connected beam remnants, and infrared MPIs. In particular, its normalisation is directly connected to the MPI cross section determined by the low-$x$ behaviour of the gluon parton-distribution function and by the consequent colour screening effects at the \pT\ cut-off, while its multiplicity distribution is more influenced by correlations within MPI in the fragmentation stage. 
  
  The methods adopted in this analysis are based on those used in the inclusive \dndeta\ (\incldNdeta) measurements of ALICE~\cite{Aamodt:2009aa,Aamodt:2010ft,Aamodt:2010pp,Adam:2015gka,Adam:2015pza}. 
  This study introduces exclusive event classes for two complementary multiplicity estimators defined in the midrapidity and in the forward regions and exploiting high-multiplicity triggers to record a large sample of events for the highest multiplicity classes. The results are provided for an event selection defined in a fully experimental way. Measurements are performed for inelastic collisions with at least one charged particle produced in $|\eta|<1$ (\inelg), corresponding to about 75\% of the total inelastic cross section~\cite{Adam:2015gka,Abelev:2012sea,Aad:2010rd,Aad:2010ac}. 
  
  \section{Experimental setup}
  
  The full description and performance of the ALICE detectors can be found elsewhere~\cite{Aamodt:2008zz,Abelev:2014ffa}. The detectors used in this analysis are briefly presented below.

  The V0 detector~\cite{Abbas:2013taa} is made of two arrays (V0A and
  V0C) of 32 scintillating counters each. The V0A is located at a distance of \unit[329]{cm} away from the interaction point (IP) along the beam direction ($z$) and it covers the pseudorapidity range $2.8 < \eta <5.1$.  The V0C is installed at $z$~=~\unit[$-88$]{cm}, covering the pseudorapidity range $-3.7 < \eta < -1.7$. Both counters cover the full azimuth. The V0 detector provides the minimum-bias and beam-gas removal trigger to ALICE. It measures the signal amplitude created by charged particles and their arrival times with a time resolution better than \unit[1]{ns}. 
  
  The Silicon Pixel Detector (SPD)~\cite{CERN-LHCC-99-012,Santoro:2009zza} is the innermost detector of ALICE. It is located inside a large solenoid that produces a homogeneous magnetic field of \unit[0.5]{T}. The SPD consists of two cylindrical layers coaxial to the beam line at radii 3.9 and \unit[7.6]{cm}. It is made of 10 million pixels distributed on 240 sensors that cover the pseudorapidity range $|\eta|<2$ for the first layer and $|\eta|<1.4$ for the second layer for particles that originate from collisions at the nominal interaction point. An enlarged pseudorapidity coverage of $|\eta|<2$ is reached using events whose primary vertex is not at zero, but within \unit[$\pm10$]{cm} from the nominal interaction point. The SPD provides a precise measurement of the position of the primary interaction vertex with a spatial resolution of on average \unit[30]{\textmu$\mathrm{m}$} in the beam direction~\cite{Santoro:2009zza,Adam:2015gka}. The multiplicity measurement of this analysis relies on the reconstruction of tracklets, which are track segments connecting hits on the two SPD layers and pointing to the primary vertex. Due to the bending in the magnetic field and multiple scattering, the reconstruction efficiency of tracklets is limited to \unit[$\pt>50$]{Mev/$c$}.   
  
  \section {Data sample and analysis}
  
  \begin{figure}[htb!]
    \centering
      \begin{subfigure}{0.48\textwidth}
          \includegraphics[width=\linewidth]{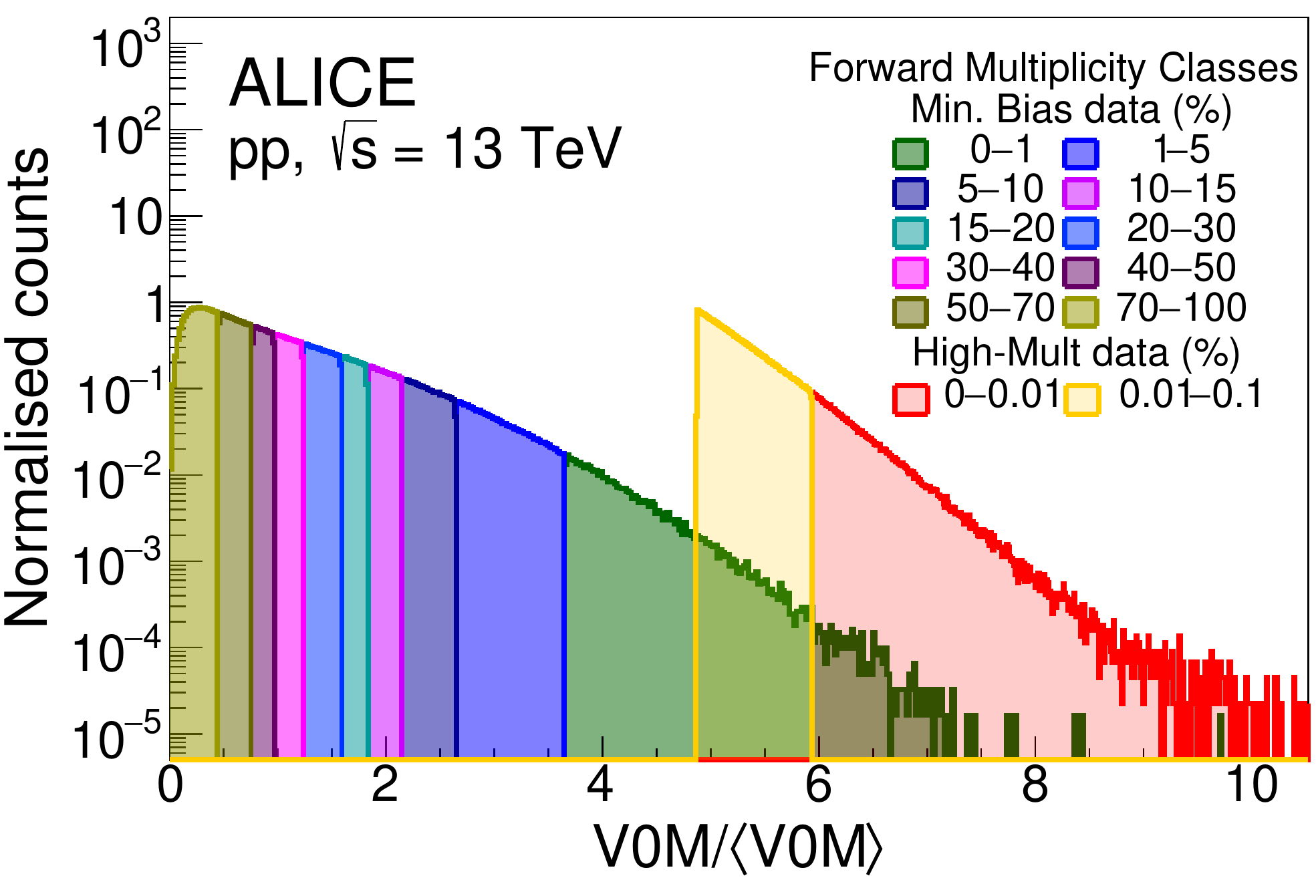}
          \caption{}
          \label{amplitude_vom}
      \end{subfigure}
      \begin{subfigure}{0.48\textwidth}
          \includegraphics[width=\linewidth]{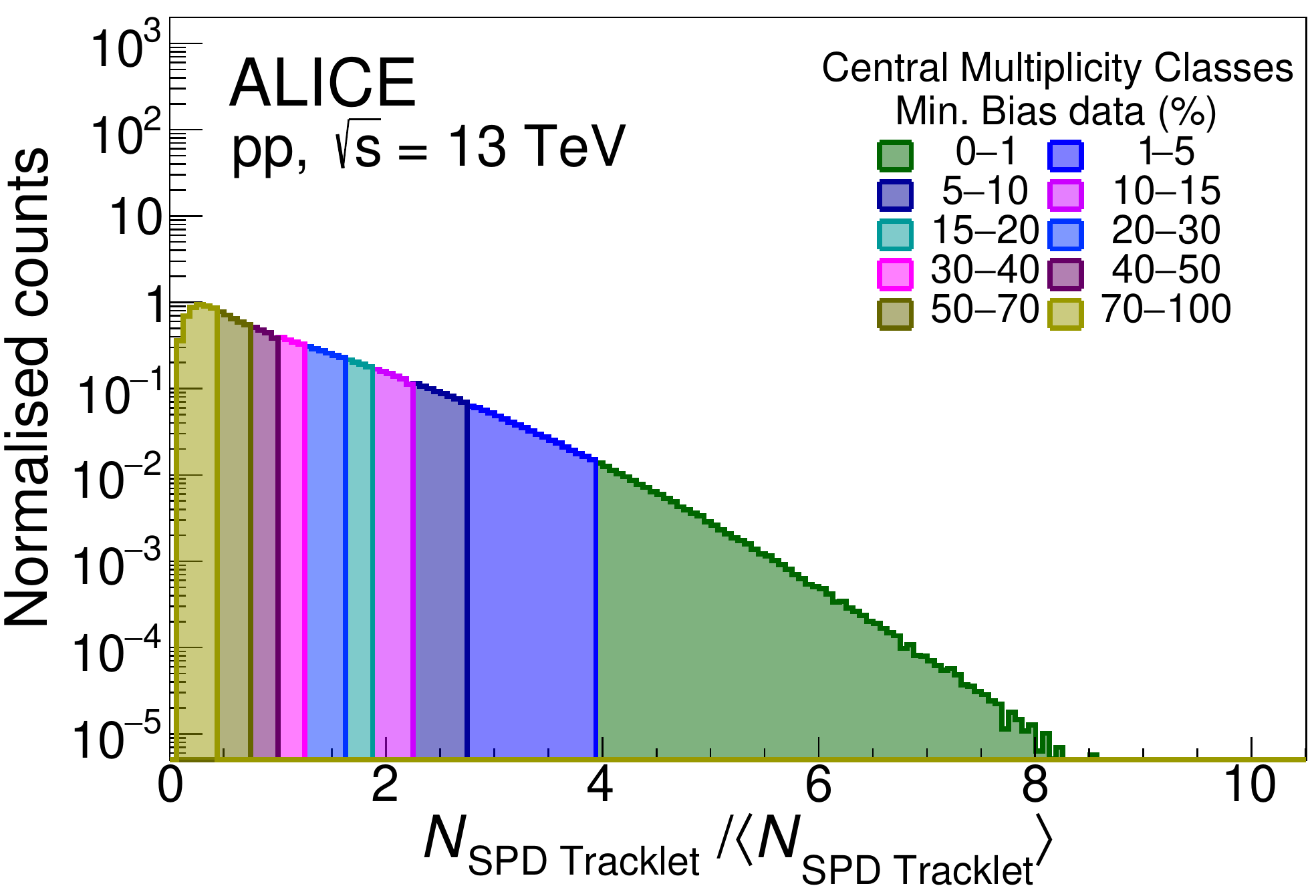}
          \caption{}
          \label{amplitude_spd}
      \end{subfigure}
      \caption{The distribution of the V0M amplitude (total energy deposition in the region $-3.7<\eta<-1.7$ and $2.8<\eta<5.1$) scaled by its average value $\langle \rm V0M \rangle$ that is used to determine the forward multiplicity classes (a) and the distribution of the total number of SPD tracklets in an event ($N_\mathrm{SPD\,\,Tracklet}$, $-2<\eta<2$) scaled by its average value $\langle N_\mathrm{SPD\,\,Tracklet} \rangle$ that is used to determine the midrapidity multiplicity classes (b) in pp collisions at $\s$~=~\unit[13]{TeV}. Note that the percentile values of the multiplicity classes are fractions of the visible cross section $\multc$ (see text for details).}
      \label{fig:amplitude}
  \end{figure}
  
  The minimum-bias pp data samples at \s~=~5.02, 7 and \unit[13]{TeV} used in this analysis correspond to the integrated luminosities $\mathcal{L}_\mathrm{int}$~=~12.4$\pm$0.3, 3.78$\pm$0.13 and \unit[0.946$\pm$0.020]{nb$^{-1}$}, respectively~\cite{ALICE-PUBLIC-2018-014,Abelev:2014ffa,ALICE-PUBLIC-2016-002}.
  The data sample at \s~=~\unit[13]{TeV} benefits from a high-multiplicity trigger that was implemented in ALICE at the beginning of the LHC Run 2. 
  
  The minimum-bias trigger (\mband) requires hits in both the V0A and V0C detectors in coincidence of a beam crossing. The contribution from diffractive interactions is minimised by requiring at least one SPD tracklet in $|\eta|<1$; the resulting data sample is called $\rm{MB}_{AND>0}$. The contamination from beam-induced background is removed by using the timing information of the V0 detectors and taking into account the correlation between tracklets and clusters in the SPD detector~\cite{Abelev:2014ffa}. The events used for the analysis are required to have a primary vertex in the fiducial region \unit[$|z|<10$]{cm}. The primary vertex is reconstructed by correlating hits in the two SPD layers. The contamination from in-bunch pile-up events is removed offline excluding events with multiple vertices reconstructed in the SPD~\cite{Adam:2015gka}. 
  The pile-up probability estimated considering the beam conditions ranges from $10^{-3}$ to $10^{-2}$. After the offline rejection, the remaining pile-up has a negligible impact on the final results. This was verified by analysing data samples separately with high and low initial pile-up contamination.

  Multiplicity classes are defined by a probability (percentile) range that is interpreted as a fractional cross section \multc, with the visible cross section in pp collisions, $\sigma_\mathrm{MB_{AND>0}}$, constituting 100\%. Percentile values for higher multiplicity collisions are close to 0\% and for lower ones close to 100\%.  
  Forward multiplicity classes are estimated by V0M, which is the sum of the energy deposition measured by the V0A and V0C scintillators. The distribution of the V0M amplitude scaled by its average value $\langle \rm V0M \rangle$ (self-normalised V0M) is shown in Fig.~\ref{amplitude_vom} for \mbandg\ pp collisions at $\sqrt{s}$ = \unit[13]{TeV}. 
  A dedicated high-multiplicity trigger is defined by the threshold $\rm V0M / \langle V0M \rangle > \sim4.9$, corresponding to $\multcs~=~0.1\%$. The SPD tracklets are used to define multiplicity classes in the midrapidity region $|\eta| < 2$. 
  The distribution of the self-normalised number of SPD tracklets for \mbandg\ pp collisions in $|\eta|<2$ is shown in Fig.~\ref{amplitude_spd}. For all the midrapidity multiplicity classes, only the minimum-bias trigger is used because the high-multiplicity trigger relying on V0M amplitudes would give an additional bias.
  The data analysis is performed by classifying $\mbandg$ data samples using the mid and forward multiplicity estimators. 
  
  \begin{table}[hbt!]
  \centering
  \footnotesize
  \begin{tabular}{@{} c|c|c|c|c|c|c @{}} 
  \multirow{3}{*}{} & \multicolumn{3}{c|}{Forward Multiplicity Estimator} & \multicolumn{3}{c}{Midrapidity Multiplicity Estimator} \\\cline{2-7}
          & \multicolumn{3}{c|}{\s~(TeV)}    & \multicolumn{3}{c}{\s~(TeV)} \\\cline{2-7}
          &  5.02 & 7 & 13  & 5.02 & 7 & 13 \\
  \hline
  $\mathrm{P}(\mathrm{MB}_\mathrm{AND_{>0}})$  & \multicolumn{3}{c|}{$\mathrm{P}(\mathrm{INEL}_\mathrm{>0})$}   & \multicolumn{3}{c}{$\mathrm{P}(\mathrm{INEL}_\mathrm{>0})$ }   \\
  (\%)   & \multicolumn{3}{c|}{(\%) }   & \multicolumn{3}{c}{(\%) }   \\
  \hline
  0--0.01 & 0--0.0091 & 0--0.0090 & 0--0.0091 \\
  0.01--0.1 &  0.0091--0.0915 & 0.0090--0.0897 & 0.0091--0.0915 \\
  0.1--0.5 & 0.0915--0.4576 & 0.0897--0.4478 & 0.0915--0.4573\\
  0.5--1 & 0.4576--0.9152 & 0.4478--0.8955 & 0.4573--0.9146 \\
  \hline
  0--1 &  0--0.9152  & 0--0.8955 & 0--0.9146 & 0--0.9095 & 0--0.8887 & 0--0.9288\\
  \hline
  1--5 & 0.9152--4.577 & 0.8955--4.478 & 0.9146--4.574 & 0.9095--4.548 & 0.8887--4.444& 0.9288--4.644 \\
  \hline
  0--5 & 0--4.577 & 0--4.478 &   0--4.574 & 0--4.548 & 0--4.444 & 0--4.644  \\
  \hline
  5--10 & 4.577--9.156 &  4.478--8.956 & 4.574--9.149 & 4.548--9.096 & 4.444--8.888 & 4.644--9.288 \\
  10--15 & 9.156--13.74 & 8.956--13.44 & 9.149--13.73 & 9.096--13.65 & 8.888--13.33 & 9.288--13.93\\
  15--20 & 13.74--18.32 & 13.44--17.92 & 13.73--18.31 & 13.65--18.20 & 13.33--17.78 & 13.93--18.58\\
  20--30 & 18.32--27.51 & 17.92--26.90 & 18.31--27.50 & 18.20--27.32 & 17.78--26.67 & 18.58--27.88\\
  30--40 & 27.51--36.76 & 26.90--35.92 & 27.50--36.75 & 27.32--36.49 & 26.67--35.59 & 27.88--37.20\\
  40--50 & 36.76--46.11 & 35.92--45.02 & 36.75--46.12 & 36.49--45.77 & 35.59--44.53 & 37.20--46.58\\
  50--70 & 46.11--65.45 & 45.02--63.66 & 46.12--65.53 & 45.77--64.91 & 44.53--62.88 & 46.58--65.82\\
  70--100 & 65.45--100 & 63.66--100 & 65.53--100 & 64.91--100 & 62.88--100 & 65.82--100\\
  
  \hline
  \end{tabular}
  \caption{Correspondence of the multiplicity classes between $\mathrm{P}(\mbandg)$ and $\mathrm{P}(\inelg)$. The trigger efficiency is estimated using PYTHIA 8 Monash 2013 ~\cite{Sjostrand:2006za,Sjostrand:2007gs,Skands:2014pea} and GEANT~3~\cite{Brun:1987ma}.}
  \label{tab:classconversion}
  \end{table}

  The multiplicity percentile intervals of the visible cross section $\mathrm{P}(\mbandg) = \multc$ can be converted to fractional intervals with respect to the \inelg\ cross section $\mathrm{P}(\inelg) = \Delta\sigma/\sigma_{\inelg}$ in pp collisions by following the conversion rule
  
  \begin{equation}
  \label{eq-triggerefficiency}
  \mathrm{P}_i(\inelg) = \frac{\mathrm{P}_i(\mbandg) / \epsilon_i }{ \sum_j \left(\mathrm{P}_j(\mbandg) / \epsilon_j\right)}\quad,
  \end{equation}
  where $i$ indicates a specific multiplicity class, $j$ runs over all multiplicity classes for a given collision energy and multiplicity estimator, and $\epsilon_i$ ($\epsilon_j$) is the \mbandg\ trigger efficiency for the \inelg\ event sample $N_{\mbandg} / N_{\inelg}$ for the $i^{\rm th}$ ($j^{\rm th}$) multiplicity class. The correspondence between $\mathrm{P}(\inelg)$ and $\mathrm{P}(\mbandg)$ is reported in Table~\ref{tab:classconversion}. In this document, multiplicity classes for the results of ALICE are represented  with $\mathrm{P}(\mbandg)$, which is a quantity defined using detector-level variables. In order to perform precise comparisons of particle-level simulations with the ALICE data, the $\mathrm{P}(\inelg)$ intervals corresponding to a given $\mathrm{P}(\mbandg)$ interval for each centre-of-mass energy and multiplicity class reported in Table~\ref{tab:classconversion} need to be used in the particle-level simulations.

  \begin{table}[hbt!]
  \centering
  \footnotesize
  \begin{tabular}{ c|c|c|c|c|c|c } 
   \multirow{3}{*}{} & \multicolumn{3}{c|}{Forward Multiplicity Estimator} & \multicolumn{3}{c}{Mid-rapidity Multiplicity Estimator} \\\cline{2-7}
        & \multicolumn{3}{c|}{\s~(TeV)}    & \multicolumn{3}{c}{\s~(TeV)} \\\cline{2-7}
        &  5.02 & 7 & 13  & 5.02 & 7 & 13 \\
   \hline
   $\mathrm{P}(\mathrm{INEL}_\mathrm{>0}) (\%)$  & \multicolumn{3}{c|}{Correction factor}    & \multicolumn{3}{c}{Correction factor}   \\
   \hline
   0--0.01 & 0.9995 & 0.9959 & 0.9842 \\
   0.01--0.1 &  0.9938 & 0.9921 & 0.9939 \\
   0.1--0.5 & 0.9934 & 0.9907 & 0.9933\\
   0.5--1 & 0.9916 & 0.9901 & 0.9915 \\
   \hline
   0--1 &  0.9927  & 0.9906 & 0.9924 & 0.9892 & 0.9853 & 0.9924\\
   \hline
   1--5 & 0.9864 & 0.9827 & 0.9855 & 0.9809 & 0.9768 & 0.9842 \\
   5--10 & 0.9768 &  0.9722 & 0.9763 & 0.9709 & 0.9634 & 0.9778 \\
   10--15 & 0.9694 & 0.9588 & 0.9667 & 0.9607 & 0.9522 & 0.9684\\
   15--20 & 0.9565 & 0.9473 & 0.9545 & 0.9516 & 0.9392 & 0.9580\\
   20--30 & 0.9455 & 0.9289 & 0.9382 & 0.9369 & 0.9210 & 0.9472\\
   30--40 & 0.9249 & 0.9072 & 0.9187 & 0.9205 & 0.8968 & 0.9290\\
   40--50 & 0.9052 & 0.8752 & 0.9003 & 0.9010 & 0.8730 & 0.9147\\
   50--70 & 0.9242 & 0.8867 & 0.8998 & 0.8962 & 0.8534 & 0.9003\\
   70--100 & 0.9716 & 0.9573 & 0.9662 & 0.9215 & 0.8897 & 0.9284\\

   \hline
   \end{tabular}
   \caption{The correction factors of $\dndeta$ from the multiplicity classes $\mathrm{P}(\inelg)$ in Table~\ref{tab:classconversion} to those of $\mathrm{P}(\inelg)$ in the leftmost column of this table. The correction factors are estimated for the generated values of $\dndeta$ using PYTHIA 8 Monash 2013~\cite{Sjostrand:2006za,Sjostrand:2007gs,Skands:2014pea}. } 
   \label{tab:inelclassconversion}

   \end{table}

  Alternatively, the values of $\dndeta$ with ALICE data for the multiplicity classes of $\mathrm{P}(\mbandg)$ in Table~\ref{tab:classconversion} can be corrected such that they correspond to the multiplicity classes of $\mathrm{P}(\inelg)$ given in the leftmost column of Table~\ref{tab:inelclassconversion}. For example, the correction factor of 0.9995 for the $\mathrm{P}(\inelg) =$ 0--0.01\% interval of the forward multiplicity estimator at $\s = 5.02$~TeV is the ratio of the generated values of $\dndeta$ between $\mathrm{P}(\inelg) =$ 0--0.01\% and 0--0.0091\% with PYTHIA 8 Monash 2013~\cite{Sjostrand:2006za,Sjostrand:2007gs,Skands:2014pea}. The data measurement of $\dndeta$ for $\mathrm{P}(\mbandg) =$ 0–0.01\% would therefore need to be multiplied by this factor in order to compare directly with a generated interval of $\mathrm{P}(\inelg) =$ 0--0.01\%.
  
  The value of \dndeta\ is obtained by correcting the number of SPD tracklets for detector acceptance as well as reconstruction and selection efficiency following the procedure developed earlier~\cite{Adam:2015gka,Adam:2015pza,Aamodt:2010cz,Adam:2015ptt,Acharya:2018hhy}. The corrections are estimated with Monte Carlo simulations based on PYTHIA 8 Monash 2013~\cite{Sjostrand:2006za,Sjostrand:2007gs,Skands:2014pea} for particle generation and GEANT 3~\cite{Brun:1987ma} for the transport of particles through the geometry of ALICE. PYTHIA 8 has a strangeness content that underestimates the data by a \pT-dependent factor, which approaches 2 around \unit[$\pT = 10$]{GeV/$c$}~\cite{Acharya:2020uxl}. The discrepancy is resolved by normalising the
  strangeness content in PYTHIA 8 to match the one in the data. This corrects \dndeta\ downward by about 1\%.
  
  \section{Systematic uncertainties}

  \begin{table}[hbt!]
  \centering
  \scriptsize
  \begin{tabular}{@{} c|c|c|c|c|c|c|c @{}} 
  \multirow{2}{*}{} & \multicolumn{7}{c}{Uncertainty (\%) at \s~=~\unit[13]{TeV}}   \\\cline{2-8}
      & \multicolumn{3}{c|}{Forward \multc} & \multicolumn{3}{c|}{Midrapidity \multc} & \multcc \\ 
  \hline
  source    &  0--0.01\% & 40--50\% & 70--100\% &  0--1\% & 40--50\% & 70--100\% & 0--100\%  \\
  \hline
  \multicolumn{8}{c}{Uncorrelated} \\
  \hline
  Trigger efficiency & neg. & 0.2 & 0.2 & neg. & 0.2 & 0.2 & 0.2 \\
  Strangeness correction & 0.7 & 0.6 & 0.5 & 0.7 & 0.6 & 0.5 & 0.5 \\
  Zero-\pT\ extrapolation & 0.7 & 0.8 & 1.0 &  0.7 & 0.9 & 1.0  & 1.0 \\
  \hline
  \multicolumn{8}{c}{Correlated} \\
  \hline
  Model dependence & neg & 0.1 & 0.1  & 0.1  & 0.1 & 0.1 & 0.1\\
  Detector acceptance and efficiency  & 0.8 & 0.7 &  0.6  &  1.8 & 2.0 & 2.8 &  0.7   \\
  Particle composition & 0.5 & 0.5 & 0.5 & 0.5 & 0.5 & 0.5 & 0.5 \\ 
  Material budget & 0.2 & 0.2 & 0.2 & 0.2 & 0.2 & 0.2 & 0.2 \\

  \hline
  \end{tabular}
  \caption{Systematic uncertainties from the highest to the lowest multiplicity class for both the mid- and forward rapidity multiplicity estimators in pp collisions at $\sqrt{s}$~=~\unit[13]{TeV}. The last column reports the effects on the inclusive \avdndeta. }
  \label{tab:systuncertainty}
  \end{table}
  
  Several sources of systematic uncertainties are investigated for this study and the estimated uncertainties are listed in Table~\ref{tab:systuncertainty}. For each multiplicity class, the systematic uncertainties related to the model used in the correction procedure (``Model dependence'') are quoted as the difference of the results using corrections obtained with two different generators before the trigger efficiency correction: PYTHIA 8 Monash 2013~\cite{Sjostrand:2006za,Sjostrand:2007gs,Skands:2014pea} and EPOS LHC~\cite{pierog2013epos,Drescher:2000ha}. The uncertainties attributed to the description of the trigger (``Trigger efficiency'') are also quoted as the difference of the simulated trigger efficiency ($N_{\rm MB_{AND>0}}/N_{\rm INEL_{>0}}$) between the two event generators.
  
  The effects of the difference in particle composition between data and Monte Carlo mostly originate from the underestimated yield related to  the weak decays of light-flavour hadrons in the simulation and are obtained with reweighting techniques (``Strangeness correction''):
  strangeness yields in the simulation are reweighted during the correction step by a factor of 2 to be compatible with the data; the factor is varied by $\pm 30\%$ based on data~\cite{Acharya:2020uxl} that covers the whole $\pT$ region, resulting in variations of the obtained \dndeta\ ranging from $\pm$0.5\% at low multiplicities to $\pm$0.7\% at the highest multiplicities. Additionally, the effect of particle-species composition (``Particle composition'') is estimated by varying, in the simulation, the relative fraction of charged kaons, protons and other particles with respect to the fixed number of charged pions by $\pm 30\%$, which covers the uncertainties in the measured particle-species composition at the LHC~\cite{Abelev:2014laa}. Relative variations of the final result are below $\pm0.5$\% in all multiplicity classes. Below \unit[50]{MeV/$c$}, the tracklet reconstruction efficiency sharply drops because of the bending in the magnetic field and to less extent due to the scattering and absorption in the detector material. To estimate the uncertainty due to the extrapolation to zero $\pT$ (``Zero-$\pT$ extrapolation''), the number of particles below \unit[50]{MeV/$c$} is varied sufficiently in the event generator by $+100$\% and $-50$\%, adopted from the previous study~\cite{Adam:2015gka}. The corresponding uncertainty is around $\pm$1\% and slightly dependent on the multiplicity class.
  
  The effect of the limited tracking acceptance and efficiency (``Detector acceptance and efficiency'') is estimated by varying the range of primary vertex selection along the beam direction (\zvtx) from \unit[$|z_\mathrm{vtx}|<10$]{cm} to the narrower \unit[$|z_\mathrm{vtx}|<7$]{cm} and broader \unit[$|z_\mathrm{vtx}|<15$]{cm}; the effect on \dndeta\ is below $\pm2$\% in all the multiplicity classes. The uncertainty due to the non-uniformity in azimuthal acceptance is studied by measuring \dndeta\ independently in three different azimuthal regions of the SPD, which are then compared with the corresponding full azimuth measurement: it varies from $\pm0.8$\% to $\pm2$\% with respect to the SPD configuration. The corresponding uncertainty is summed in quadrature for that in ``Detector acceptance and efficiency''. The material budget in the ALICE central barrel is known to a precision of about 5$\%$~\cite{Abelev:2014ffa}. The corresponding systematic uncertainty on \dndeta (``Material budget''), obtained by varying the material budget in the simulation, is estimated to be about $\pm0.2$\%. 
  
  Variations for the particle-species composition, material budget, tracking acceptance, and efficiency correction produce a change in the measurement that behaves the same across energies and multiplicity classes. The corresponding systematic uncertainties are considered then as correlated. Conversely, variations on the correction for the contribution of strangeness particles, the trigger efficiency, and the extrapolation to zero-$\pT$ affect each energy and multiplicity class differently, so these contributions are considered as uncorrelated.

  \section {Results}
  
  \begin{figure}[hbt!]
    \centering
   \includegraphics[width=0.8\linewidth]{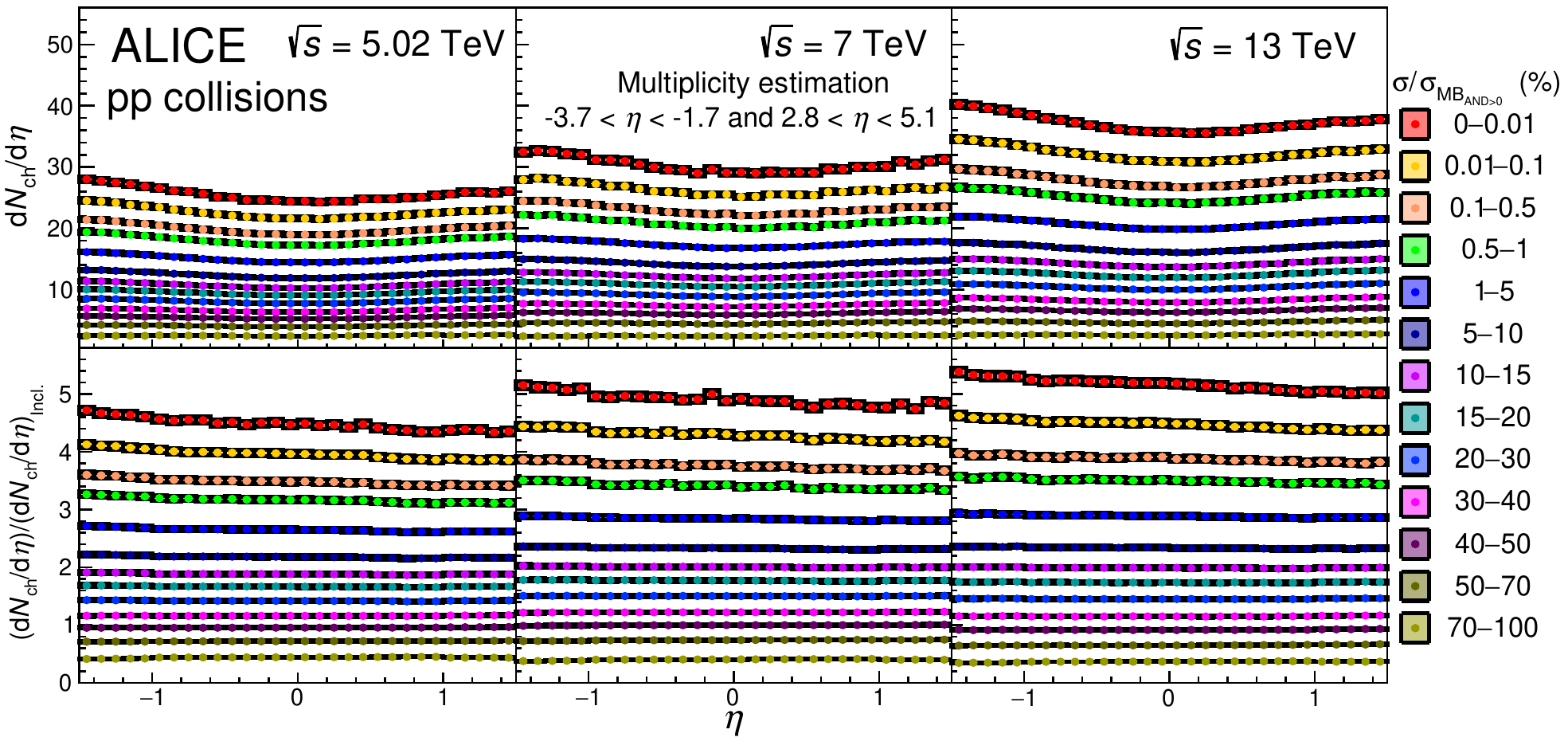}
    \caption{Charged-particle pseudorapidity density (upper panels)
    and the same scaled by $1/\left(\dndeta\right)_\mathrm{incl.}$ (lower panels) for the 0--0.01 to 70--100\% multiplicity classes
    measured with the forward multiplicity estimator ($-3.7<\eta<-1.7$ and $2.8<\eta<5.1$) in pp collisions at \s~=~5.02, 7 and \unit[13]{TeV}. Correlated and uncorrelated systematic uncertainties are summed in quadrature in the upper panels and shown as boxes. Correlated systematic uncertainties are cancelled out in the lower panels.}
    \label{dndetav0m}
  \end{figure}
  
  \begin{figure}[hbt!]
   \centering
          \includegraphics[width=0.9\linewidth]{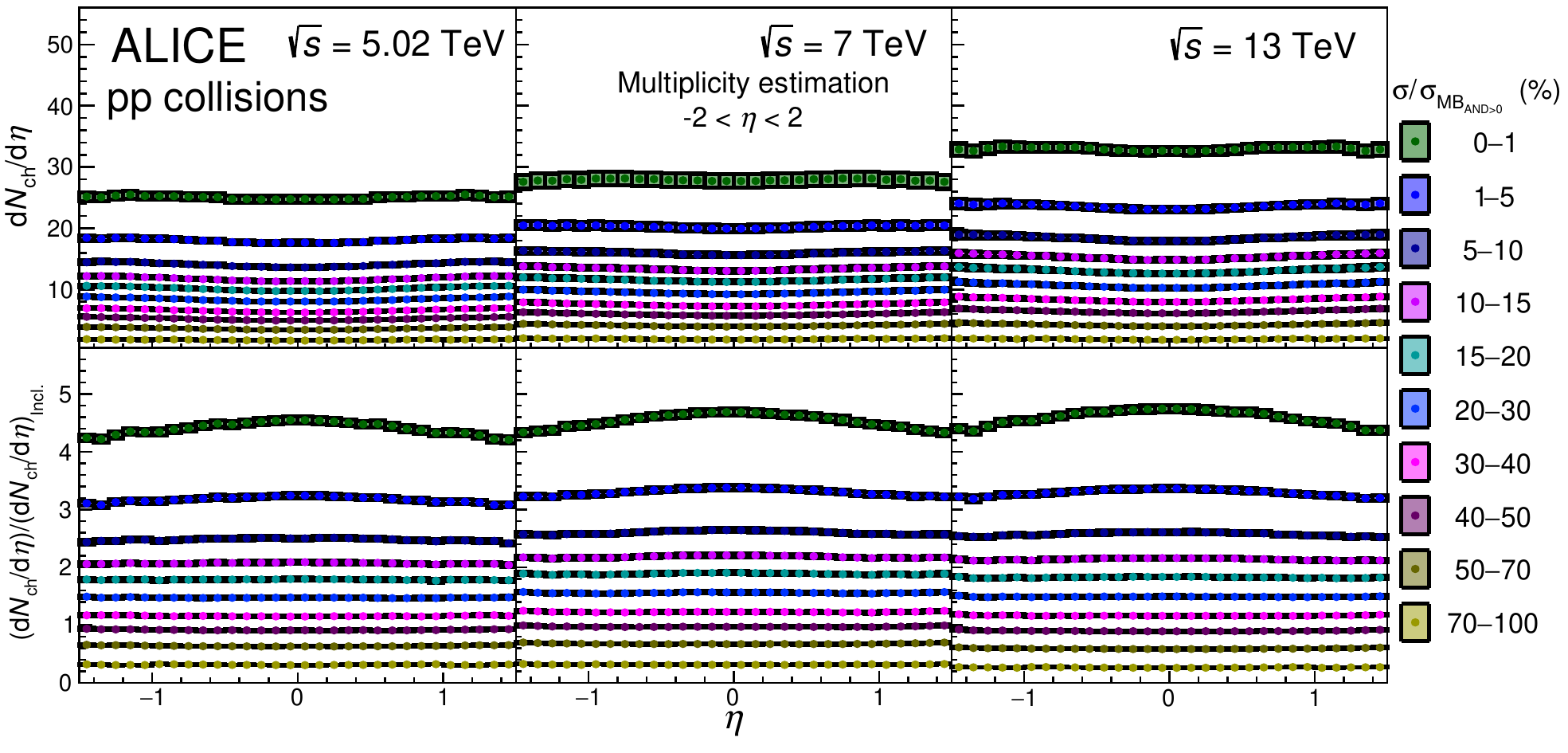}
      \caption{Charged-particle pseudorapidity density
       (upper panels) and the same scaled by $1/\left(\dndeta\right)_\mathrm{incl.}$ (lower panels) for the 0--1 to 70--100\% multiplicity classes measured with the midrapidity multiplicity estimator ($-2<\eta<2$) in pp collisions at \s~=~5.02, 7 and \unit[13]{TeV}. Correlated and uncorrelated systematic uncertainties are summed in quadrature in the upper panels and shown as boxes. Correlated systematic uncertainties are cancelled out in the lower panels.}
    \label{dndeta_central}
  \end{figure}

  The \dndeta\ measurements at \s~=~5.02, 7 and
  \unit[13]{TeV} for different classes of the forward multiplicity estimators are reported in Fig.~\ref{dndetav0m}; in the upper panels in absolute scale and in the lower panels, normalised to the inclusive \dndeta (\incldNdeta, \dndeta for 0--100\%). As shown in the lower panels of Fig.~\ref{dndetav0m}, the pseudorapidity densities for the highest multiplicity classes (0--0.01\%) are around 5 times larger than those of the inclusive ones for the three different collision energies. The asymmetry of the \dndeta\ distributions for the forward multiplicity classes is due to the asymmetric pseudorapidity acceptance of the V0 detector. This effect is more pronounced for the highest multiplicity classes.

  The upper panels in Fig.~\ref{dndeta_central} show the \dndeta\ measurements at \s~=~5.02, 7 and \unit[13]{TeV} for different multiplicity classes defined by the midrapidity multiplicity estimator. The shapes of the pseudorapidity distributions of primary charged particles are different when compared with those obtained with the forward multiplicity estimator. The midrapidity multiplicity estimator is defined in a symmetric pseudorapidity region ($-2<\eta<2$) and clearly gives rise to autocorrelations as it includes the region where the pseudorapidity distributions are measured ($-1.5<\eta<1.5$). As shown in the lower panels of Fig.~\ref{dndeta_central}, for the three different collision energies, the pseudorapidity densities for the highest multiplicity classes (0--1\%) are around 4--5 times larger than those of the inclusive ones, with the highest enhancement observed at midrapidity ($\eta = 0$).

  \begin{figure}[hbt!]
    \centering
      \includegraphics[width=0.9\linewidth]{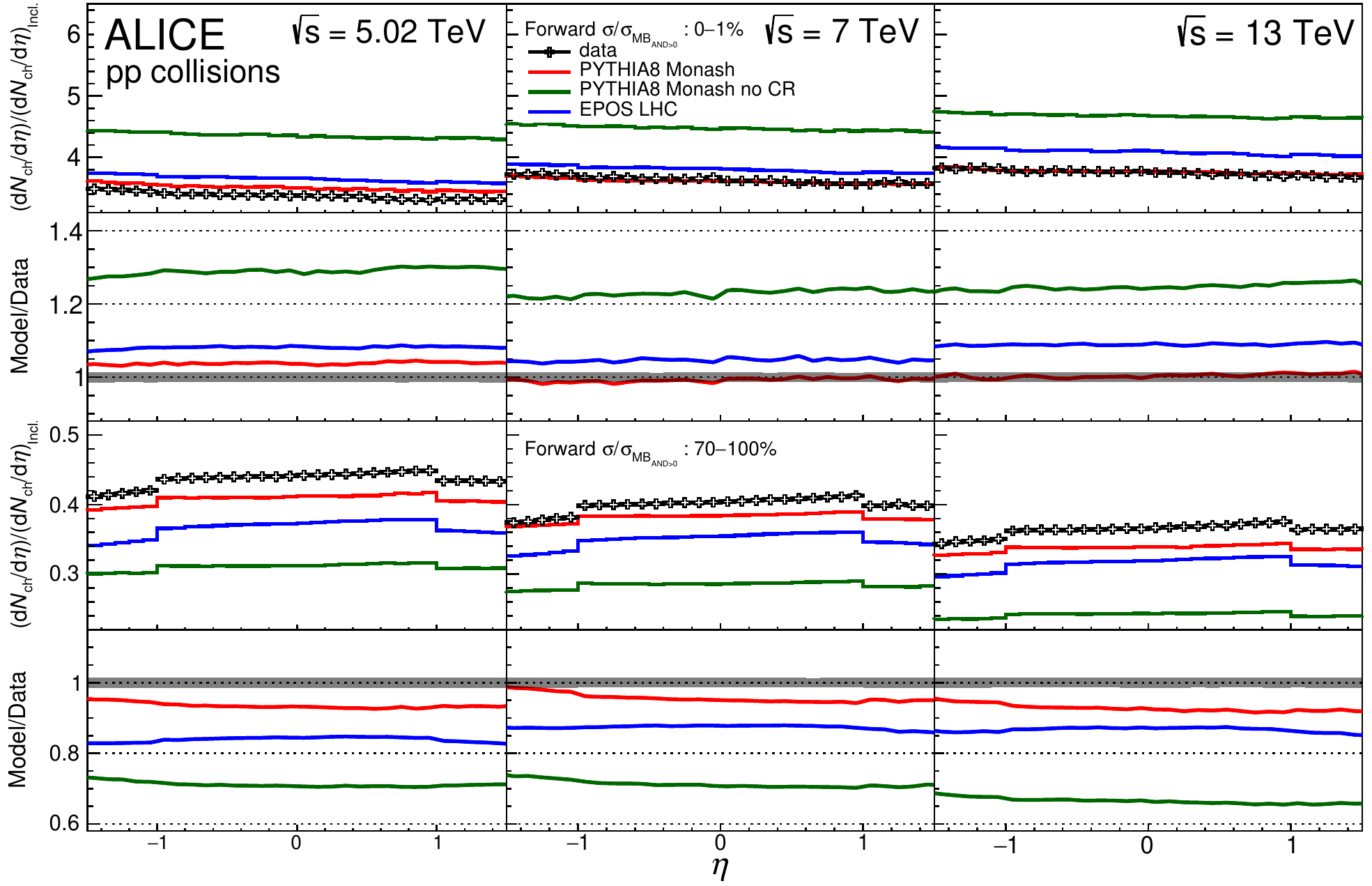}
  \caption{The panels in the first and third row show the normalised pseudorapidity density distributions of charged particles in pp
  collisions at \s~=~5.02, 7 and \unit[13]{TeV} compared with different models for the 0--1\% and 70--100\% multiplicity classes by the forward rapidity multiplicity estimator, respectively. The panels in the second and fourth row report the corresponding model/data ratio. Note that the multiplicity classes of the models correspond to \multcc, which is slightly different from the \multc\ of the ALICE data.}
    \label{modelcomparison-forward}
  \end{figure}

  The measurements are compared with the predictions from PYTHIA 8 Monash 2013~\cite{Sjostrand:2006za,Sjostrand:2007gs,Skands:2014pea} with and without CR and the ones from EPOS LHC~\cite{pierog2013epos,Drescher:2000ha}. The effect of CR can be explored with PYTHIA 8 Monash 2013 by switching the effect on and off. EPOS LHC describes the collectivity effect in high multiplicity pp collisions differently with a hydrodynamic evolution of the core with a high-energy density that is formed by many colour string fields. The multiplicity classes of the models are estimated for generated charged particles in the same geometrical acceptances of the forward rapidity ($-3.7<\eta<-1.7$ and $2.8<\eta<5.1$) and midrapidity ($|\eta|<2$) multiplicity estimators and the percentile value of the multiplicity class is calibrated for generated \inelg\ events. Figure~\ref{modelcomparison-forward} reports the comparison of the data with these models for the 0--1\% and 70--100\% classes by the forward multiplicity estimator. PYTHIA 8 Monash 2013, implementing CR in the string fragmentation process, describes the data  within 5\% for all the centre-of-mass energies for the 0--1\% multiplicity class. For the 70--100\% class, PYTHIA 8 underestimates the data by up to 10\%. When switching off CR, while keeping all the other model parameters stable, PYTHIA 8 overestimates (underestimates) the data by about 30\% for the 0--1\% (70--100\%) multiplicity class. EPOS LHC, which incorporates a collective flow-like description of the core, describes the data within 20\% for both forward multiplicity classes. EPOS LHC also overestimates (underestimates) the data for the 0--1\% (70--100\%) multiplicity class like PYTHIA 8 Monash 2013. For the two classes, PYTHIA 8 describes the data better than EPOS LHC. 
  
  \begin{figure}[hbt!]
    \centering
      \includegraphics[width=0.9\linewidth]{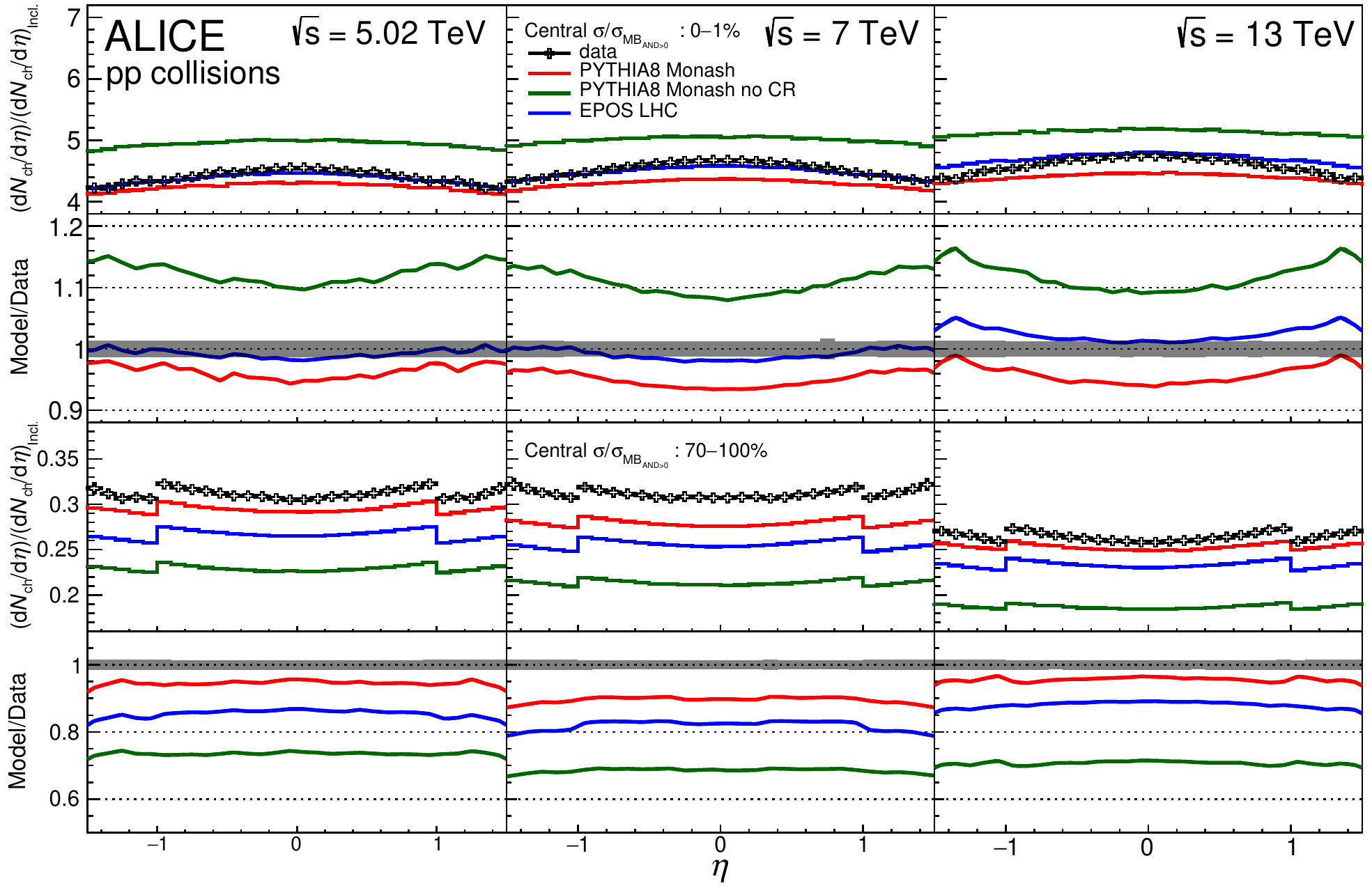}
  \caption{The panels in the first and third row show the normalised pseudorapidity density distributions of charged particles in pp
  collisions at \s~=~5.02, 7 and \unit[13]{TeV} compared with different models for the 0--1\% and 70--100\% multiplicity classes by the midrapidity multiplicity estimator, respectively. The panels in the second and fourth row report the corresponding model/data ratio. The multiplicity classes of the models correspond to \multcc, which is slightly different from the \multc\ of the ALICE data.}
    \label{modelcomparison-midrapidity}
  \end{figure}
  
  Figure~\ref{modelcomparison-midrapidity} shows the comparison according to the data with these models for the 0--1\% and 70--100\% classes by the midrapidity multiplicity estimator. EPOS LHC describes the data  within 5\% for all the centre-of-mass energies for the 0--1\% multiplicity class. For the 70--100\% class, EPOS LHC underestimates the data by up to 20\%. PYTHIA 8 reproduces the data within 5\% for all centre-of-mass energies for the 0--1\% multiplicity class, but it is not as good as EPOS LHC in the 0--1\% multiplicity class. For the 70--100\% class, PYTHIA 8 describes the data within 10\% and it is better than those of EPOS LHC. When switching off CR, PYTHIA 8 overestimates (underestimates) the data by about 15\% (30\%) for the 0--1\% (70--100\%) multiplicity class.

  \begin{table}[t]
  \centering
  \scriptsize
  \scalebox{.95}{
  \begin{tabular}{@{} c|c|c|c|c|c|c @{}} 
  \multirow{3}{*}{} & \multicolumn{3}{c|}{Forward Multiplicity Estimator} & \multicolumn{3}{c}{Midrapidity Multiplicity Estimator} \\\cline{2-7}
          & \multicolumn{3}{c|}{\s~(TeV)}    & \multicolumn{3}{c}{\s~(TeV)} \\\cline{2-7}
          &  5.02 & 7 & 13  & 5.02 & 7 & 13 \\
  \hline
  \multc\  & \multicolumn{6}{c}{\avdndeta$\pm$uncorrelated systematic uncertainty$\pm$correlated systematic uncertainty}  \\
  \hline
  0--0.01\% & 24.53$\pm$0.23$\pm$0.31 & 29.13$\pm$0.25$\pm$0.44 & 35.82$\pm$0.33$\pm$0.33 \\\cline{1-4}
  0.01--0.05\% &  22.42$\pm$0.21$\pm$0.23 & 26.27$\pm$0.23$\pm$0.30 & 32.21$\pm$0.29$\pm$0.29 \\
  0.05--0.1\% &  21.14$\pm$0.20$\pm$0.22 & 24.70$\pm$0.22$\pm$0.25 & 30.13$\pm$0.27$\pm$0.27 \\\cline{1-4}
  0.01--0.1\% &   21.71$\pm$0.20$\pm$0.21 & 25.40$\pm$0.22$\pm$0.26 & 31.05$\pm$0.28$\pm$0.28 \\\cline{1-4}
  0.1--0.5\% & 19.08$\pm$0.18$\pm$0.17 & 22.24$\pm$0.19$\pm$0.20 & 26.91$\pm$0.24$\pm$0.27\\
  0.5--1\% & 17.34$\pm$0.16$\pm$0.15 & 20.11$\pm$0.18$\pm$0.18 & 24.26$\pm$0.22$\pm$0.26 \\
  \hline
  0--1\% &  18.50$\pm$0.17$\pm$0.16  & 21.55$\pm$0.19$\pm$0.19 &  26.01$\pm$0.24$\pm$0.24 & 24.74$\pm$0.23$\pm$0.52 &27.80$\pm$0.24$\pm$1.08 & 32.70$\pm$0.29$\pm$0.60\\
  1--5\% & 14.51$\pm$0.14$\pm$0.12 & 16.85$\pm$0.15$\pm$0.11 & 19.99$\pm$0.18$\pm$0.16 & 17.66$\pm$0.16$\pm$0.29 & 19.97$\pm$0.18$\pm$0.56 & 23.21$\pm$0.21$\pm$0.40 \\
  \hline
  0--5\% & 15.30$\pm$0.14$\pm$0.13 & 17.80$\pm$0.16$\pm$0.11 & 21.18$\pm$0.19$\pm$0.17 & 19.08$\pm$0.18$\pm$0.4  & 21.46$\pm$0.20$\pm$0.59 & 25.08$\pm$0.20$\pm$0.38 \\
  \hline
  5--10\% & 11.93$\pm$0.11$\pm$0.10 &  13.82$\pm$0.12$\pm$0.09 & 16.18$\pm$0.15$\pm$0.13 & 13.71$\pm$0.13$\pm$0.19 & 15.64$\pm$0.14$\pm$0.35 & 18.03$\pm$0.17$\pm$0.33 \\
  10--15\% & 10.30$\pm$0.10$\pm$0.09 & 11.89$\pm$0.11$\pm$0.07 & 13.78$\pm$0.13$\pm$0.12 & 11.40$\pm$0.11$\pm$0.13 & 13.06$\pm$0.12$\pm$0.26 & 14.94$\pm$0.14$\pm$0.27\\
  15--20\% & 9.12$\pm$0.09$\pm$0.08 & 10.49$\pm$0.10$\pm$0.06 & 12.01$\pm$0.11$\pm$0.11 & 9.81$\pm$0.09$\pm$0.11 & 11.27$\pm$0.10$\pm$0.22 & 12.69$\pm$0.12$\pm$0.24\\
  20--30\% & 7.76$\pm$0.08$\pm$0.07 & 8.90$\pm$0.08$\pm$0.05 & 10.03$\pm$0.10$\pm$0.09 & 8.07$\pm$0.08$\pm$0.08 & 9.29$\pm$0.09$\pm$0.18 & 10.33$\pm$0.10$\pm$0.20\\
  30--40\% & 6.34$\pm$0.06$\pm$0.06 & 7.24$\pm$0.07$\pm$0.04 & 7.95$\pm$0.08$\pm$0.07 & 6.30$\pm$0.06$\pm$0.06 & 7.30$\pm$0.07$\pm$0.15 & 8.03$\pm$0.08$\pm$0.16\\
  40--50\% & 5.22$\pm$0.05$\pm$0.05 & 5.92$\pm$0.06$\pm$0.03 & 6.32$\pm$0.06$\pm$0.06 & 4.98$\pm$0.05$\pm$0.05 & 5.76$\pm$0.06$\pm$0.12 &  6.18$\pm$0.06$\pm$0.12\\
  50--70\% & 3.94$\pm$0.04$\pm$0.04 & 4.39$\pm$0.04$\pm$0.02 & 4.49$\pm$0.05$\pm$0.04 & 3.45$\pm$0.04$\pm$0.04 & 3.97$\pm$0.04$\pm$0.09 & 4.05$\pm$0.04$\pm$0.08\\
  70--100\% & 2.42$\pm$0.02$\pm$0.03 & 2.40$\pm$0.02$\pm$0.01 & 2.54$\pm$0.03$\pm$0.02 & 1.69$\pm$0.02$\pm$0.05 & 1.84$\pm$0.02$\pm$0.06 & 1.80$\pm$0.02$\pm$0.05\\
  \hline
  0--100\% &  5.48$\pm$0.05$\pm$0.05 & 5.94$\pm$0.06$\pm$0.03  & 6.93$\pm$0.07$\pm$0.06  & 5.48$\pm$0.05$\pm$0.05  &  5.94$\pm$0.06$\pm$0.03  &  6.93$\pm$0.07$\pm$0.06 \\
  \hline
  \end{tabular}
  }
  \caption{Values of \avdndeta\ for different multiplicity classes defined by the mid and forward multiplicity estimators in pp collisions at \s~=~5.02 to \unit[13]{TeV}. Statistical uncertainties are negligible.}
  \label{tab:dndeta}
  \end{table}

  The value of \avdndeta\ is determined by integrating \dndeta\ in $|\eta|<0.5$.
  Table~\ref{tab:dndeta} shows the values of \avdndeta\ for different mid- and forward rapidity multiplicity classes in pp collisions at \s~=~5.02, 7 and \unit[13]{TeV}. The autocorrelation effect for the midrapidity estimator results in larger values of \avdndeta\ in the highest  multiplicity classes and in smaller ones  for the lowest multiplicity classes compared with those with the forward multiplicity estimator.
  
  \begin{figure}[hbt!]
   \centering
   \includegraphics[width=0.7\linewidth]{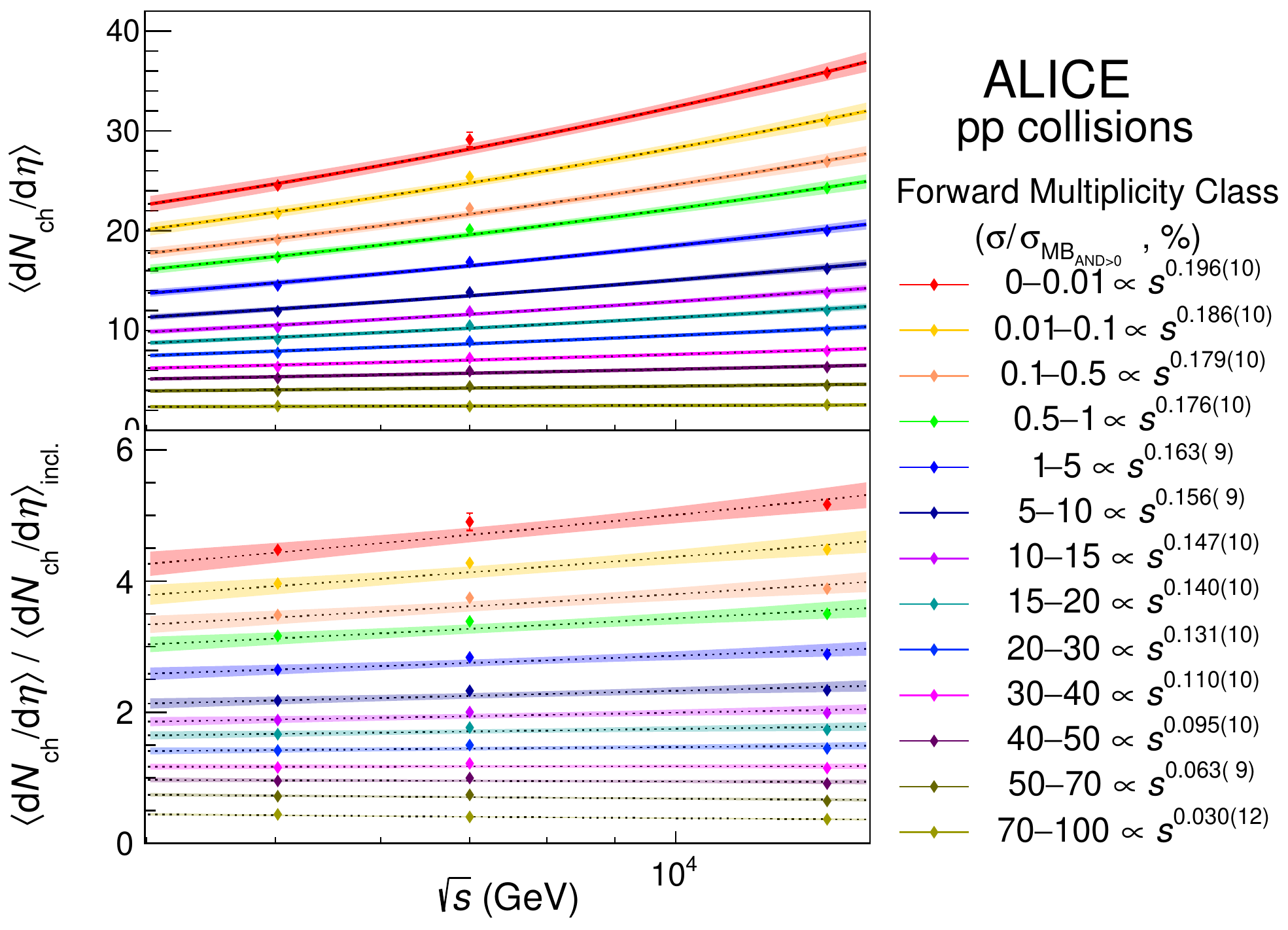}
   \caption{Energy dependence of \avdndeta\ (upper)
   and \avdndeta\ scaled by the inclusive \dndeta\ (lower)
   for the multiplicity classes by the forward multiplicity estimator in pp collisions. Lines show fits with a power-law function $s^\alpha$. Corresponding bands indicate one standard deviation of the fit. Exponents and corresponding uncertainties of the fit are listed in the legend.}
   \label{energy}
  \end{figure}
  
  The energy dependence of \avdndeta\ for the multiplicity classes defined by the forward multiplicity estimator is shown in the upper panel of Fig.~\ref{energy}.  
  The LHC measurements for the \avdndeta\ can be directly compared with the ones from the NAL Bubble Chamber (p$\overline{\mathrm p}$)~\cite{Whitmore:1973ri}, ISR (pp)~\cite{Thome:1977ky}, UA1 (p$\overline{\mathrm p}$)~\cite{Albajar:1989an}, UA5 (p$\overline{\mathrm p}$)~\cite{Alner:1986xu}, CDF (p$\overline{\mathrm p}$)~\cite{Abe:1989td}, STAR (pp)~\cite{Abelev:2008ab} and PHOBOS (pp)~\cite{Nouicer:2004ke}.  A phenomenological power-law fit $s^\alpha$ describes the centre-of-mass energy ($\s$) evolution of these measurements for Non-Single Diffractive (NSD), INEL and \inelg\ events up to LHC energies~\cite{Adam:2015gka}. 
    
  Such a fit is performed practically for the values of Table~\ref{tab:dndeta} in different multiplicity classes to describe the dependence of \avdndeta\ on the centre-of-mass energy. Corresponding exponents are shown in the legend of Fig.~\ref{energy}. The average pseudorapidity density at midrapidity as a function of centre-of-mass energy increases rapidly for higher multiplicity classes. The lower panel of Fig.~\ref{energy} shows \avdndeta\ normalised to its inclusive value denoted as \navdndeta\ for the forward multiplicity classes. The steeper increasing trend of \navdndeta\ observed for higher multiplicity classes may arise from the increase of the MPI cross sections with the centre-of-mass energy~\cite{Adam:2015gka}.  
  
  \begin{table}[hbt!]
  \scriptsize
  \centering
  \begin{tabular} {c | c || c | c | c } 
  $\mathrm{P}(\mathrm{MB}_\mathrm{AND_{>0}})$ (\%) & ALICE & $\mathrm{P}(\mathrm{INEL}_\mathrm{>0})$ (\%) & PYTHIA 8 Monash 2013 & EPOS LHC   \\
  \hline
  0--0.01 & 0.196(10) & 0--0.01 & 0.158(7) & 0.170(8)\\
  0.01--0.1 & 0.186(10) & 0.01--0.1 & 0.166(4) & 0.165(4)\\
  0.1--0.5 & 0.179(10) & 0.1--0.5 & 0.160(3) & 0.172(2)\\
  0.5--1 & 0.176(10) & 0.5--1 & 0.166(3) & 0.171(1)\\
  1--5& 0.163(9) & 1--5 & 0.166(1) & 0.169(2)\\
  5--10 & 0.156(9) & 5--10 & 0.163(0) & 0.163(1)\\
  10--15 & 0.147(10) & 10--15 & 0.155(1) & 0.151(2)\\
  15--20 & 0.140(10) & 15--20 & 0.150(0) & 0.138(1)\\
  20--30 & 0.131(10) & 20--30 & 0.138(1) & 0.117(1)\\
  30--40 & 0.110(10) & 30--40 & 0.124(1) & 0.093(1) \\
  40--50 & 0.095(10) & 40--50 & 0.120(1) & 0.068(3)\\
  50--70 & 0.063(9) & 50--70 & 0.078(1) & 0.034(1)\\
  70--100 & 0.030(12) & 70--100 & 0.024(0) & 0.029(1)\\
  
  \hline
  \end{tabular}
  \caption{The comparison of the exponents $\alpha$ of the power-law fit ($s^\alpha$) for \avdndeta\ as a function of centre-of-mass energy for each forward multiplicity class between the data and models. Note that the forward multiplicity classes correspond to $\mathrm{P}(\mathrm{MB}_\mathrm{AND_{>0}})$ for ALICE data, while $\mathrm{P}(\mathrm{INEL}_\mathrm{>0})$ for models.}
  \label{tab:energyfit}
  \end{table}
  
  The exponent values $\alpha$ of the power-law fit ($s^{\alpha}$) of ALICE data in Fig.~\ref{energy} are compared with those of PYTHIA 8 and EPOS LHC in Table~\ref{tab:energyfit} for the forward multiplicity classes. The multiplicity classes represent $\mathrm{P}(\mathrm{MB}_\mathrm{AND_{>0}})$ for ALICE data, while $\mathrm{P}(\mathrm{INEL}_\mathrm{>0})$ for the models. Overall, the energy dependence of $\avdndeta$ of the data for different multiplicity classes is not described well by the models. Also, the exponent values of the fit for the models fail to describe the steeper behaviour of the energy dependence of $\avdndeta$ that is measured in data with increasing multiplicities for the highest multiplicity classes. This suggests more tuning is needed to constrain models for the energy dependence of charged particle production with respect to different multiplicity classes.

  \section{Conclusions}
  
  The energy and multiplicity dependence of the charged-particle pseudorapidity density \dndeta\ and the average charged-particle pseudorapidity density \avdndeta\ in proton--proton (pp) collisions at \s~=~5.02, 7 and \unit[13]{TeV} are measured. The yields of charged particles in the 0--1\% and 0--0.01\% multiplicity classes for the mid- and forward rapidity multiplicity estimators, respectively, are up to about a factor of 5 higher with respect to the inclusive measurements for all investigated centre-of-mass energies.
  The results from the multiplicity-dependent analysis presented for both the mid- and forward rapidity multiplicity estimators in ALICE can be used as an input for improving our understanding of Multiple Parton Interactions (MPIs) implemented in Monte Carlo models. Most of the results are described well by PYTHIA 8 with the Monash tune and by EPOS LHC. The effects of the colour reconnection (CR) is found to be important to constrain MPIs and describe the scale of the pseudorapidity density as a function of multiplicity for both the mid and forward multiplicity estimators as seen by the expected values for PYTHIA 8 with and without CR. The results can be used for further studies as a function of multiplicity estimated at mid- or forward rapidity in pp collisions.


\newenvironment{acknowledgement}{\relax}{\relax}
\begin{acknowledgement}
\section*{Acknowledgements}

The ALICE Collaboration would like to thank all its engineers and technicians for their invaluable contributions to the construction of the experiment and the CERN accelerator teams for the outstanding performance of the LHC complex.
The ALICE Collaboration gratefully acknowledges the resources and support provided by all Grid centres and the Worldwide LHC Computing Grid (WLCG) collaboration.
The ALICE Collaboration acknowledges the following funding agencies for their support in building and running the ALICE detector:
A. I. Alikhanyan National Science Laboratory (Yerevan Physics Institute) Foundation (ANSL), State Committee of Science and World Federation of Scientists (WFS), Armenia;
Austrian Academy of Sciences, Austrian Science Fund (FWF): [M 2467-N36] and Nationalstiftung f\"{u}r Forschung, Technologie und Entwicklung, Austria;
Ministry of Communications and High Technologies, National Nuclear Research Center, Azerbaijan;
Conselho Nacional de Desenvolvimento Cient\'{\i}fico e Tecnol\'{o}gico (CNPq), Financiadora de Estudos e Projetos (Finep), Funda\c{c}\~{a}o de Amparo \`{a} Pesquisa do Estado de S\~{a}o Paulo (FAPESP) and Universidade Federal do Rio Grande do Sul (UFRGS), Brazil;
Ministry of Education of China (MOEC) , Ministry of Science \& Technology of China (MSTC) and National Natural Science Foundation of China (NSFC), China;
Ministry of Science and Education and Croatian Science Foundation, Croatia;
Centro de Aplicaciones Tecnol\'{o}gicas y Desarrollo Nuclear (CEADEN), Cubaenerg\'{\i}a, Cuba;
Ministry of Education, Youth and Sports of the Czech Republic, Czech Republic;
The Danish Council for Independent Research | Natural Sciences, the VILLUM FONDEN and Danish National Research Foundation (DNRF), Denmark;
Helsinki Institute of Physics (HIP), Finland;
Commissariat \`{a} l'Energie Atomique (CEA) and Institut National de Physique Nucl\'{e}aire et de Physique des Particules (IN2P3) and Centre National de la Recherche Scientifique (CNRS), France;
Bundesministerium f\"{u}r Bildung und Forschung (BMBF) and GSI Helmholtzzentrum f\"{u}r Schwerionenforschung GmbH, Germany;
General Secretariat for Research and Technology, Ministry of Education, Research and Religions, Greece;
National Research, Development and Innovation Office, Hungary;
Department of Atomic Energy Government of India (DAE), Department of Science and Technology, Government of India (DST), University Grants Commission, Government of India (UGC) and Council of Scientific and Industrial Research (CSIR), India;
Indonesian Institute of Science, Indonesia;
Centro Fermi - Museo Storico della Fisica e Centro Studi e Ricerche Enrico Fermi and Istituto Nazionale di Fisica Nucleare (INFN), Italy;
Institute for Innovative Science and Technology , Nagasaki Institute of Applied Science (IIST), Japanese Ministry of Education, Culture, Sports, Science and Technology (MEXT) and Japan Society for the Promotion of Science (JSPS) KAKENHI, Japan;
Consejo Nacional de Ciencia (CONACYT) y Tecnolog\'{i}a, through Fondo de Cooperaci\'{o}n Internacional en Ciencia y Tecnolog\'{i}a (FONCICYT) and Direcci\'{o}n General de Asuntos del Personal Academico (DGAPA), Mexico;
Nederlandse Organisatie voor Wetenschappelijk Onderzoek (NWO), Netherlands;
The Research Council of Norway, Norway;
Commission on Science and Technology for Sustainable Development in the South (COMSATS), Pakistan;
Pontificia Universidad Cat\'{o}lica del Per\'{u}, Peru;
Ministry of Science and Higher Education, National Science Centre and WUT ID-UB, Poland;
Korea Institute of Science and Technology Information and National Research Foundation of Korea (NRF), Republic of Korea;
Ministry of Education and Scientific Research, Institute of Atomic Physics and Ministry of Research and Innovation and Institute of Atomic Physics, Romania;
Joint Institute for Nuclear Research (JINR), Ministry of Education and Science of the Russian Federation, National Research Centre Kurchatov Institute, Russian Science Foundation and Russian Foundation for Basic Research, Russia;
Ministry of Education, Science, Research and Sport of the Slovak Republic, Slovakia;
National Research Foundation of South Africa, South Africa;
Swedish Research Council (VR) and Knut \& Alice Wallenberg Foundation (KAW), Sweden;
European Organization for Nuclear Research, Switzerland;
Suranaree University of Technology (SUT), National Science and Technology Development Agency (NSDTA) and Office of the Higher Education Commission under NRU project of Thailand, Thailand;
Turkish Atomic Energy Agency (TAEK), Turkey;
National Academy of  Sciences of Ukraine, Ukraine;
Science and Technology Facilities Council (STFC), United Kingdom;
National Science Foundation of the United States of America (NSF) and United States Department of Energy, Office of Nuclear Physics (DOE NP), United States of America. 
\end{acknowledgement}

\bibliographystyle{utphys}
\bibliography{dndeta}

\newpage
\appendix

%
%

\section{The ALICE Collaboration}
\label{app:collab}

\begingroup
\small
\begin{flushleft}
S.~Acharya\Irefn{org142}\And 
D.~Adamov\'{a}\Irefn{org96}\And 
A.~Adler\Irefn{org74}\And 
J.~Adolfsson\Irefn{org81}\And 
M.M.~Aggarwal\Irefn{org101}\And 
S.~Agha\Irefn{org14}\And 
G.~Aglieri Rinella\Irefn{org34}\And 
M.~Agnello\Irefn{org30}\And 
N.~Agrawal\Irefn{org54}\textsuperscript{,}\Irefn{org10}\And 
Z.~Ahammed\Irefn{org142}
% ...
\renewcommand\labelenumi{\textsuperscript{\theenumi}~}

\section*{Affiliation notes}
\renewcommand\theenumi{\roman{enumi}}
\begin{Authlist}
\item \Adef{orgI} Deceased
\item \Adef{orgII} Italian National Agency for New Technologies, Energy and Sustainable Economic Development (ENEA), Bologna, Italy
\item \Adef{orgIII} Dipartimento DET del Politecnico di Torino, Turin, Italy
\item \Adef{orgIV} M.V. Lomonosov Moscow State University, D.V. Skobeltsyn Institute of Nuclear, Physics, Moscow, Russia
\item \Adef{orgV} Department of Applied Physics, Aligarh Muslim University, Aligarh, India
\item \Adef{orgVI} Institute of Theoretical Physics, University of Wroclaw, Poland
\end{Authlist}

\section*{Collaboration Institutes}
\renewcommand\theenumi{\arabic{enumi}~}
\begin{Authlist}
\item \Idef{org1} A.I. Alikhanyan National Science Laboratory (Yerevan Physics Institute) Foundation, Yerevan, Armenia
\item \Idef{org2} Bogolyubov Institute for Theoretical Physics, National Academy of Sciences of Ukraine, Kiev, Ukraine
\item \Idef{org3} Bose Institute, Department of Physics  and Centre for Astroparticle Physics and Space Science (CAPSS), Kolkata, India
\item \Idef{org4} Budker Institute for Nuclear Physics, Novosibirsk, Russia
\item \Idef{org5} California Polytechnic State University, San Luis Obispo, California, United States
\item \Idef{org6} Central China Normal University, Wuhan, China
\item \Idef{org7} Centre de Calcul de l'IN2P3, Villeurbanne, Lyon, France
\item \Idef{org8} Centro de Aplicaciones Tecnol\'{o}gicas y Desarrollo Nuclear (CEADEN), Havana, Cuba
\item \Idef{org9} Centro de Investigaci\'{o}n y de Estudios Avanzados (CINVESTAV), Mexico City and M\'{e}rida, Mexico
\item \Idef{org10} Centro Fermi - Museo Storico della Fisica e Centro Studi e Ricerche ``Enrico Fermi', Rome, Italy
\item \Idef{org11} Chicago State University, Chicago, Illinois, United States
\item \Idef{org12} China Institute of Atomic Energy, Beijing, China
\item \Idef{org13} Comenius University Bratislava, Faculty of Mathematics, Physics and Informatics, Bratislava, Slovakia
\item \Idef{org14} COMSATS University Islamabad, Islamabad, Pakistan
\item \Idef{org15} Creighton University, Omaha, Nebraska, United States
\item \Idef{org16} Department of Physics, Aligarh Muslim University, Aligarh, India
\item \Idef{org17} Department of Physics, Pusan National University, Pusan, Republic of Korea
\item \Idef{org18} Department of Physics, Sejong University, Seoul, Republic of Korea
\item \Idef{org19} Department of Physics, University of California, Berkeley, California, United States
\item \Idef{org20} Department of Physics, University of Oslo, Oslo, Norway
\item \Idef{org21} Department of Physics and Technology, University of Bergen, Bergen, Norway
\item \Idef{org22} Dipartimento di Fisica dell'Universit\`{a} 'La Sapienza' and Sezione INFN, Rome, Italy
\item \Idef{org23} Dipartimento di Fisica dell'Universit\`{a} and Sezione INFN, Cagliari, Italy
\item \Idef{org24} Dipartimento di Fisica dell'Universit\`{a} and Sezione INFN, Trieste, Italy
\item \Idef{org25} Dipartimento di Fisica dell'Universit\`{a} and Sezione INFN, Turin, Italy
\item \Idef{org26} Dipartimento di Fisica e Astronomia dell'Universit\`{a} and Sezione INFN, Bologna, Italy
\item \Idef{org27} Dipartimento di Fisica e Astronomia dell'Universit\`{a} and Sezione INFN, Catania, Italy
\item \Idef{org28} Dipartimento di Fisica e Astronomia dell'Universit\`{a} and Sezione INFN, Padova, Italy
\item \Idef{org29} Dipartimento di Fisica `E.R.~Caianiello' dell'Universit\`{a} and Gruppo Collegato INFN, Salerno, Italy
\item \Idef{org30} Dipartimento DISAT del Politecnico and Sezione INFN, Turin, Italy
\item \Idef{org31} Dipartimento di Scienze e Innovazione Tecnologica dell'Universit\`{a} del Piemonte Orientale and INFN Sezione di Torino, Alessandria, Italy
\item \Idef{org32} Dipartimento di Scienze MIFT, Universit\`{a} di Messina, Messina, Italy
\item \Idef{org33} Dipartimento Interateneo di Fisica `M.~Merlin' and Sezione INFN, Bari, Italy
\item \Idef{org34} European Organization for Nuclear Research (CERN), Geneva, Switzerland
\item \Idef{org35} Faculty of Electrical Engineering, Mechanical Engineering and Naval Architecture, University of Split, Split, Croatia
\item \Idef{org36} Faculty of Engineering and Science, Western Norway University of Applied Sciences, Bergen, Norway
\item \Idef{org37} Faculty of Nuclear Sciences and Physical Engineering, Czech Technical University in Prague, Prague, Czech Republic
\item \Idef{org38} Faculty of Science, P.J.~\v{S}af\'{a}rik University, Ko\v{s}ice, Slovakia
\item \Idef{org39} Frankfurt Institute for Advanced Studies, Johann Wolfgang Goethe-Universit\"{a}t Frankfurt, Frankfurt, Germany
\item \Idef{org40} Fudan University, Shanghai, China
\item \Idef{org41} Gangneung-Wonju National University, Gangneung, Republic of Korea
\item \Idef{org42} Gauhati University, Department of Physics, Guwahati, India
\item \Idef{org43} Helmholtz-Institut f\"{u}r Strahlen- und Kernphysik, Rheinische Friedrich-Wilhelms-Universit\"{a}t Bonn, Bonn, Germany
\item \Idef{org44} Helsinki Institute of Physics (HIP), Helsinki, Finland
\item \Idef{org45} High Energy Physics Group,  Universidad Aut\'{o}noma de Puebla, Puebla, Mexico
\item \Idef{org46} Hiroshima University, Hiroshima, Japan
\item \Idef{org47} Hochschule Worms, Zentrum  f\"{u}r Technologietransfer und Telekommunikation (ZTT), Worms, Germany
\item \Idef{org48} Horia Hulubei National Institute of Physics and Nuclear Engineering, Bucharest, Romania
\item \Idef{org49} Indian Institute of Technology Bombay (IIT), Mumbai, India
\item \Idef{org50} Indian Institute of Technology Indore, Indore, India
\item \Idef{org51} Indonesian Institute of Sciences, Jakarta, Indonesia
\item \Idef{org52} INFN, Laboratori Nazionali di Frascati, Frascati, Italy
\item \Idef{org53} INFN, Sezione di Bari, Bari, Italy
\item \Idef{org54} INFN, Sezione di Bologna, Bologna, Italy
\item \Idef{org55} INFN, Sezione di Cagliari, Cagliari, Italy
\item \Idef{org56} INFN, Sezione di Catania, Catania, Italy
\item \Idef{org57} INFN, Sezione di Padova, Padova, Italy
\item \Idef{org58} INFN, Sezione di Roma, Rome, Italy
\item \Idef{org59} INFN, Sezione di Torino, Turin, Italy
\item \Idef{org60} INFN, Sezione di Trieste, Trieste, Italy
\item \Idef{org61} Inha University, Incheon, Republic of Korea
\item \Idef{org62} Institute for Gravitational and Subatomic Physics (GRASP), Utrecht University/Nikhef, Utrecht, Netherlands
\item \Idef{org63} Institute for Nuclear Research, Academy of Sciences, Moscow, Russia
\item \Idef{org64} Institute of Experimental Physics, Slovak Academy of Sciences, Ko\v{s}ice, Slovakia
\item \Idef{org65} Institute of Physics, Homi Bhabha National Institute, Bhubaneswar, India
\item \Idef{org66} Institute of Physics of the Czech Academy of Sciences, Prague, Czech Republic
\item \Idef{org67} Institute of Space Science (ISS), Bucharest, Romania
\item \Idef{org68} Institut f\"{u}r Kernphysik, Johann Wolfgang Goethe-Universit\"{a}t Frankfurt, Frankfurt, Germany
\item \Idef{org69} Instituto de Ciencias Nucleares, Universidad Nacional Aut\'{o}noma de M\'{e}xico, Mexico City, Mexico
\item \Idef{org70} Instituto de F\'{i}sica, Universidade Federal do Rio Grande do Sul (UFRGS), Porto Alegre, Brazil
\item \Idef{org71} Instituto de F\'{\i}sica, Universidad Nacional Aut\'{o}noma de M\'{e}xico, Mexico City, Mexico
\item \Idef{org72} iThemba LABS, National Research Foundation, Somerset West, South Africa
\item \Idef{org73} Jeonbuk National University, Jeonju, Republic of Korea
\item \Idef{org74} Johann-Wolfgang-Goethe Universit\"{a}t Frankfurt Institut f\"{u}r Informatik, Fachbereich Informatik und Mathematik, Frankfurt, Germany
\item \Idef{org75} Joint Institute for Nuclear Research (JINR), Dubna, Russia
\item \Idef{org76} Korea Institute of Science and Technology Information, Daejeon, Republic of Korea
\item \Idef{org77} KTO Karatay University, Konya, Turkey
\item \Idef{org78} Laboratoire de Physique des 2 Infinis, Ir\`{e}ne Joliot-Curie, Orsay, France
\item \Idef{org79} Laboratoire de Physique Subatomique et de Cosmologie, Universit\'{e} Grenoble-Alpes, CNRS-IN2P3, Grenoble, France
\item \Idef{org80} Lawrence Berkeley National Laboratory, Berkeley, California, United States
\item \Idef{org81} Lund University Department of Physics, Division of Particle Physics, Lund, Sweden
\item \Idef{org82} Moscow Institute for Physics and Technology, Moscow, Russia
\item \Idef{org83} Nagasaki Institute of Applied Science, Nagasaki, Japan
\item \Idef{org84} Nara Women{'}s University (NWU), Nara, Japan
\item \Idef{org85} National and Kapodistrian University of Athens, School of Science, Department of Physics , Athens, Greece
\item \Idef{org86} National Centre for Nuclear Research, Warsaw, Poland
\item \Idef{org87} National Institute of Science Education and Research, Homi Bhabha National Institute, Jatni, India
\item \Idef{org88} National Nuclear Research Center, Baku, Azerbaijan
\item \Idef{org89} National Research Centre Kurchatov Institute, Moscow, Russia
\item \Idef{org90} Niels Bohr Institute, University of Copenhagen, Copenhagen, Denmark
\item \Idef{org91} Nikhef, National institute for subatomic physics, Amsterdam, Netherlands
\item \Idef{org92} NRC Kurchatov Institute IHEP, Protvino, Russia
\item \Idef{org93} NRC \guillemotleft Kurchatov\guillemotright  Institute - ITEP, Moscow, Russia
\item \Idef{org94} NRNU Moscow Engineering Physics Institute, Moscow, Russia
\item \Idef{org95} Nuclear Physics Group, STFC Daresbury Laboratory, Daresbury, United Kingdom
\item \Idef{org96} Nuclear Physics Institute of the Czech Academy of Sciences, \v{R}e\v{z} u Prahy, Czech Republic
\item \Idef{org97} Oak Ridge National Laboratory, Oak Ridge, Tennessee, United States
\item \Idef{org98} Ohio State University, Columbus, Ohio, United States
\item \Idef{org99} Petersburg Nuclear Physics Institute, Gatchina, Russia
\item \Idef{org100} Physics department, Faculty of science, University of Zagreb, Zagreb, Croatia
\item \Idef{org101} Physics Department, Panjab University, Chandigarh, India
\item \Idef{org102} Physics Department, University of Jammu, Jammu, India
\item \Idef{org103} Physics Department, University of Rajasthan, Jaipur, India
\item \Idef{org104} Physikalisches Institut, Eberhard-Karls-Universit\"{a}t T\"{u}bingen, T\"{u}bingen, Germany
\item \Idef{org105} Physikalisches Institut, Ruprecht-Karls-Universit\"{a}t Heidelberg, Heidelberg, Germany
\item \Idef{org106} Physik Department, Technische Universit\"{a}t M\"{u}nchen, Munich, Germany
\item \Idef{org107} Politecnico di Bari and Sezione INFN, Bari, Italy
\item \Idef{org108} Research Division and ExtreMe Matter Institute EMMI, GSI Helmholtzzentrum f\"ur Schwerionenforschung GmbH, Darmstadt, Germany
\item \Idef{org109} Rudjer Bo\v{s}kovi\'{c} Institute, Zagreb, Croatia
\item \Idef{org110} Russian Federal Nuclear Center (VNIIEF), Sarov, Russia
\item \Idef{org111} Saha Institute of Nuclear Physics, Homi Bhabha National Institute, Kolkata, India
\item \Idef{org112} School of Physics and Astronomy, University of Birmingham, Birmingham, United Kingdom
\item \Idef{org113} Secci\'{o}n F\'{\i}sica, Departamento de Ciencias, Pontificia Universidad Cat\'{o}lica del Per\'{u}, Lima, Peru
\item \Idef{org114} St. Petersburg State University, St. Petersburg, Russia
\item \Idef{org115} Stefan Meyer Institut f\"{u}r Subatomare Physik (SMI), Vienna, Austria
\item \Idef{org116} SUBATECH, IMT Atlantique, Universit\'{e} de Nantes, CNRS-IN2P3, Nantes, France
\item \Idef{org117} Suranaree University of Technology, Nakhon Ratchasima, Thailand
\item \Idef{org118} Technical University of Ko\v{s}ice, Ko\v{s}ice, Slovakia
\item \Idef{org119} The Henryk Niewodniczanski Institute of Nuclear Physics, Polish Academy of Sciences, Cracow, Poland
\item \Idef{org120} The University of Texas at Austin, Austin, Texas, United States
\item \Idef{org121} Universidad Aut\'{o}noma de Sinaloa, Culiac\'{a}n, Mexico
\item \Idef{org122} Universidade de S\~{a}o Paulo (USP), S\~{a}o Paulo, Brazil
\item \Idef{org123} Universidade Estadual de Campinas (UNICAMP), Campinas, Brazil
\item \Idef{org124} Universidade Federal do ABC, Santo Andre, Brazil
\item \Idef{org125} University of Cape Town, Cape Town, South Africa
\item \Idef{org126} University of Houston, Houston, Texas, United States
\item \Idef{org127} University of Jyv\"{a}skyl\"{a}, Jyv\"{a}skyl\"{a}, Finland
\item \Idef{org128} University of Liverpool, Liverpool, United Kingdom
\item \Idef{org129} University of Science and Technology of China, Hefei, China
\item \Idef{org130} University of South-Eastern Norway, Tonsberg, Norway
\item \Idef{org131} University of Tennessee, Knoxville, Tennessee, United States
\item \Idef{org132} University of the Witwatersrand, Johannesburg, South Africa
\item \Idef{org133} University of Tokyo, Tokyo, Japan
\item \Idef{org134} University of Tsukuba, Tsukuba, Japan
\item \Idef{org135} Universit\'{e} Clermont Auvergne, CNRS/IN2P3, LPC, Clermont-Ferrand, France
\item \Idef{org136} Universit\'{e} de Lyon, Universit\'{e} Lyon 1, CNRS/IN2P3, IPN-Lyon, Villeurbanne, Lyon, France
\item \Idef{org137} Universit\'{e} de Strasbourg, CNRS, IPHC UMR 7178, F-67000 Strasbourg, France, Strasbourg, France
\item \Idef{org138} Universit\'{e} Paris-Saclay Centre d'Etudes de Saclay (CEA), IRFU, D\'{e}partment de Physique Nucl\'{e}aire (DPhN), Saclay, France
\item \Idef{org139} Universit\`{a} degli Studi di Foggia, Foggia, Italy
\item \Idef{org140} Universit\`{a} degli Studi di Pavia and Sezione INFN, Pavia, Italy
\item \Idef{org141} Universit\`{a} di Brescia and Sezione INFN, Brescia, Italy
\item \Idef{org142} Variable Energy Cyclotron Centre, Homi Bhabha National Institute, Kolkata, India
\item \Idef{org143} Warsaw University of Technology, Warsaw, Poland
\item \Idef{org144} Wayne State University, Detroit, Michigan, United States
\item \Idef{org145} Westf\"{a}lische Wilhelms-Universit\"{a}t M\"{u}nster, Institut f\"{u}r Kernphysik, M\"{u}nster, Germany
\item \Idef{org146} Wigner Research Centre for Physics, Budapest, Hungary
\item \Idef{org147} Yale University, New Haven, Connecticut, United States
\item \Idef{org148} Yonsei University, Seoul, Republic of Korea
\end{Authlist}
\endgroup


\end{document}